\documentclass[12pt]{article}

\usepackage{amssymb}
\usepackage[dvips]{graphicx}

\newcommand {\beq}{\begin{equation}}
\newcommand {\eeq}{\end{equation}}
\newcommand {\beqa}{\begin{eqnarray}}
\newcommand {\eeqa}{\end{eqnarray}}
\newcommand {\n}{\nonumber \\}
\newcommand {\tr}{\mbox{tr}}

\def\pa{\partial}
\renewcommand{\theequation}{\thesection.\arabic{equation}}

\begin{document}
\setlength{\oddsidemargin}{0cm}
\setlength{\baselineskip}{7mm}

\begin{titlepage}
\renewcommand{\thefootnote}{\fnsymbol{footnote}}
\begin{normalsize}
\begin{flushright}
\begin{tabular}{l}
MIT-CTP-3321 \\
OU-HET 423 \\
November 2002
\end{tabular}
\end{flushright}
  \end{normalsize}

~~\\

\vspace*{0cm}
    \begin{Large}
       \begin{center}
         {Born-Infeld Action from Supergravity}
       \end{center}
    \end{Large}
\vspace{1cm}

\begin{center}
           Matsuo S{\sc ato}$^{1)}$\footnote
            {
e-mail address : 
machan@het.phys.sci.osaka-u.ac.jp}
           {\sc and}
           Asato T{\sc suchiya}$^{1),2)}$\footnote
           {
e-mail address : asato@lns.mit.edu}\\
      \vspace{1cm}
       
        $^{1)}$ {\it Department of Physics, Graduate School of  
                     Science}\\
               {\it Osaka University, Toyonaka, Osaka 560-0043, Japan}\\
      \vspace{0.5cm}
                    
        $^{2)}$ {\it Center for Theoretical Physics}\\
                {\it Laboratory for Nuclear Science
                     and Department of Physics}\\
                {\it Massachusetts Institute of Technology,
                     Cambridge, MA 02139, USA}
\end{center}

\hspace{5cm}

\begin{abstract}
\noindent
We show that the Born-Infeld action with the Wess-Zumino terms 
for the Ramond-Ramond fields, which is the D3-brane effective action,
is a solution to the Hamilton-Jacobi (H-J) equation of type IIB supergravity. 
Adopting the radial coordinate as time, 
we develop the ADM formalism for type IIB supergravity 
reduced on $S^5$ and derive the H-J equation, which
is the classical limit of the Wheeler-De Witt equation and whose solutions
are classical on-shell actions.
The solution to the H-J equation reproduces the on-shell actions for
the supergravity solution of a stack of D3-branes in a $B_2$ field 
and the near-horizon limit of this supergravity solution, 
which is conjectured to be 
dual to noncommutative Yang Mills and reduces to $AdS_5 \times S^5$ in
the commutative limit.  Our D3-brane effective action is
that of a probe D3-brane, and the radial time corresponds to the vacuum
expectation value of the Higgs field in the dual Yang Mills. 
Our findings can be applied to the study of the holographic renormalization
group.
\end{abstract}
\vfill
\end{titlepage}
\vfil\eject

\setcounter{footnote}{0}

\section{Introduction}
\setcounter{equation}{0}
\renewcommand{\thefootnote}{\arabic{footnote}} 
Recent studies of D-branes have revealed many aspects of
the connections between gauge theories
and gravities (string theories). In particular, 
the AdS/CFT correspondence \cite{Maldacena,GKP-W}
was originally based on an observation that the dynamics of D3-branes
can be described under some circumstances both by super Yang Mills and 
by type IIB supergravity. First, the low velocity dynamics of 
$N$ D3-branes that are located almost on top each other
can be described by super Yang Mills, since in this case the higher excited 
modes of open strings can be ignored \cite{DKPS}. Second, the region near the
horizon of the $N$ D3-branes, whose geometry is $AdS_5 \times S^5$,
is described well by supergravity
when the curvature radius 
$R(=(4\pi g_{s}N)^{\frac{1}{4}}
=(4\pi g_{YM}^2 N)^{\frac{1}{4}})$ of $AdS_5$ is sufficiently large. 
Thus the open-closed string duality leads to a conjecture that 
${\cal N}=4$ super Yang Mills
with large 't Hooft coupling is dual to type IIB supergravity on 
$AdS_5 \times S^5$.

Although the correspondence between ${\cal N}=4$ super Yang Mills and type IIB
supergravity on $AdS_5 \times S^5$ has been tested mainly 
at the conformally invariant point \cite{AGMOO-DF}, 
the above consideration motivates us
to conjecture that it is also valid in the Coulomb branch \cite{DT,Das}.
Indeed, on the one hand,
the effective action of the D3-brane probing the $N$ D3-branes,
which should take the form of the Born-Infeld action \cite{Tseytlin}
on the $AdS_5$ background,
is determined only by the broken conformal invariance \cite{Maldacena}.
On the other hand,
the effective action of ${\cal N}=4$ super Yang Mills in which the $SU(N+1)$
gauge symmetry is broken to $U(1) \times SU(N)$  due to the vacuum
expectation value of the Higgs field is conjectured to take the form of
the Born-Infeld action on the $AdS_5$ background in the large 
't Hooft coupling limit \cite{BPT}.

The effective action of the probe D3-brane should be obtained in principle
by calculating (the logarithm of) the transition amplitude 
between the vacuum and the boundary 
state representing the probe D3-brane on the $AdS_5 \times S^5$ 
background in type IIB
superstring. This has not yet been accomplished, because there does not yet
exist a quantized theory of type IIB superstring 
on such a background. However, the above argument 
suggests that one can obtain
the effective action of the probe D3-brane by calculating the classical 
on-shell action in type IIB supergravity, which is the classical 
counterpart of the transition amplitude.

In this paper, we show that this is indeed the case, at least for the `flat'
probe D3-brane. Here `flat' means that we do not consider
fluctuations transverse to the world-volume. The formalism suitable
for this purpose is that of the Hamilton-Jacobi (H-J)
equation in type IIB supergravity, which is the classical limit of
the Wheeler-De Witt equation and whose solutions are 
classical on-shell actions.
We reduce type IIB supergravity on $S^5$, keeping
the anti-symmetric tensor field and the Ramond-Ramond (R-R) fields, and obtain
a five-dimensional gravity. Adopting the radial coordinate as time,
we develop the ADM formalism for
this five-dimensional gravity and derive the H-J equation.
We solve the equation under the condition that the fields be
constant on fixed-time surfaces.
We show that the Born-Infeld action with the Wess-Zumino terms for
the R-R fields is one of the solutions to the H-J equation.
In general, the H-J equation has infinitely many solutions.
Our solution to the H-J equation
is the on-shell action for various near-horizon geometries
of many D3-branes. In fact, the on-shell action for the supergravity solution
representing the near-horizon limit of 
a stack of D3-branes in a $B_2$ field \cite{HI-MR}, which is conjectured to be
dual to noncommutative Yang Mills and reduces to $AdS_5 \times S^5$ 
in the commutative limit,
is reproduced by the solution to the H-J equation.
It is conjectured that the solution to the H-J equation also includes 
the on-shell actions for general fluctuations around 
this supergravity solution. The solution to the H-J equation
is the effective action of a probe
D3-brane located in the backgrounds of the near-horizon geometries. 
The radial time
corresponds to the position of the probe D3-brane and 
the vacuum expectation value of the Higgs field in the dual Yang Mills. 
Moreover, the solution to the H-J equation
also reproduces the on-shell action for the 
supergravity solution of a stack of D3-branes in a $B_2$ field 
without the near-horizon limit. 
It is relevant to investigate
whether this result for the region outside the near-horizon
is universal or accidental and due to
the special case in which only the `flat' D3-brane is considered. 
This result should be related to the fact that the 
Laplacian for the transverse parts in the geometry generated by a stack of
D3-branes is proportional to the flat space Laplacian \cite{DT} (see also
Ref.\cite{Yoneya}). The supersymmetry should also be essential in this result,
since the R-R fields play crucial roles in our calculation, and therefore 
a counterpart to the nonrenormalization theorem in the
dual Yang Mills should hold in supergravity.
We can generalize our
analysis to the cases of general D$p$-branes.

Our results clarify a relation 
between the effective action in super Yang Mills in the
Coulomb branch and
the on-shell action in supergravity.
They also lead to
the question of whether the effective action in noncommutative Yang Mills 
or in super Yang Mills in different dimensions 
takes the form of the Born-Infeld action. 
If indeed it does take this form, this provides
strong evidence of the duality of noncommutative Yang Mills or 
super Yang Mills in different dimensions and 
supergravities on curved backgrounds. 
We hope to address this problem
elsewhere. In general, when one studies 
the gauge/string duality based
on the D-brane picture, it is necessary to work first 
in a region of coupling strengths in which the supergravity 
approximation is valid,
since the quantization of strings on curved backgrounds has not yet
been developed
well. Therefore, we believe that the study presented in this paper 
represents a prototype for 
approaches to this problem.

Another motivation of our work is to understand more general
holographic renormalization group flows. 
The authors of Ref.\cite{dBVV} analyzed the H-J equations around 
general AdS backgrounds and derived the holographic renormalization group
equation for the dual gauge theories that are perturbed by the operators
dual to the scalar fields in gravities. (For further developments, see 
Refs.\cite{FMS,Corley,KM,NOZ,NOO,MM}.) 
Our solution to the H-J equation can be interpreted as a potential
that gives the renormalization group flows generated by the perturbations of
the operators dual to the tensor fields. Furthermore, our study is expected
to be useful for understanding the
holographic renormalization group of noncommutative Yang Mills.

The organization of the paper is as follows. In section 2, we perform 
a reduction of type IIB supergravity on $S^5$ and obtain a five-dimensional
gravity. The self-duality condition for the R-R 5-form is treated carefully.
In section 3, we develop a canonical formalism for the 
five-dimensional gravity based on the ADM decomposition and derive the
H-J equation. In section 4, we show that 
the D3-brane effective action
is a solution to the H-J equation. In section 5, after reviewing
the supergravity solution representing a stack of D3-branes 
in a $B_2$ field and its near-horizon limit, 
we show that the on-shell actions for
the supergravity solution and its near-horizon limit
are reproduced by the solution to the H-J equation. In section 6,
we show that the solution to the H-J equation obtained in section 4 is the effective action
of a probe D3-brane.
Section 7 is devoted to summary and
discussion. In particular, we comment on the extension of our results to the
cases of general D$p$-branes. The equations of motion 
in type IIB supergravity are listed
in appendix A. Some useful formulae are gathered in appendix B. In appendix C,
we elucidate the meaning of the momentum constraint and the Gauss law 
constraints obtained in section 3.

\vspace{1cm}

\section{Reduction of type IIB supergravity on $S^5$}
\setcounter{equation}{0}
In this section, we reduce type IIB supergravity on $S^5$ and obtain
a five-dimensional gravity. In this paper, we drop the fermionic degrees
of freedom consistently. The bosonic part of type IIB supergravity is given
by
\beqa
&&I_{10}=\frac{1}{2\kappa_{10}^{\:2}} \int d^{10} X \sqrt{-G}
\left[ e^{-2\Phi} \left(R_G+4\pa_M \Phi \pa^M \Phi-\frac{1}{2}|H_3|^2\right) 
\right. \n
&& \left. \qquad \qquad \qquad \qquad \qquad \qquad
-\frac{1}{2}|F_1|^2-\frac{1}{2}|\tilde{F}_3|^2
-\frac{1}{4}|\tilde{F}_5|^2 \right] \n
&& \qquad \;\; +\frac{1}{4\kappa_{10}^{\:2}} \int C_4 \wedge H_3 \wedge F_3,
\label{10Daction}
\eeqa
where
\beqa
&&H_3=d B_2, \;\; F_{p+2}=d C_{p+1} \;\; (p=-1,1,3), \n
&&\tilde{F}_3=F_3+C_0 \wedge H_3, \n
&&\tilde{F}_{5}=F_{5}+C_2 \wedge H_3,
\eeqa
the $X^M$ $(M=0,\cdots,9)$ are ten-dimensional coordinates,
and $C_{p+1}$ is the R-R $(p+1)$-form. 
In the above equations, 
$|K_q|^2=\frac{1}{q!}G^{M_1 N_1} \cdots G^{M_q N_q} 
K_{M_1 \cdots M_q} K_{N_1 \cdots N_q}$ for a
$q$-form $K_q$. One must also impose the self-duality condition
\beq
\ast \tilde{F_5}=\tilde{F_5}
\label{self-duality}
\eeq
on the equations of motion derived from the above action. For completeness,
we list all the equations of motion and the self-duality condition in
type IIB supergravity explicitly in appendix A.

In order to perform a reduction on $S^5$, we split the ten-dimensional
coordinates $X^M$ into two parts, as $X^M=(\xi^{\alpha},\; \theta_i)
\;\;(\alpha=0,\cdots,4, \;\; i=1,\cdots,5)$, where the $\xi^{\alpha}$ are
five-dimensional coordinates and the $\theta_i$ parametrize
$S^5$, and we adopt the following ansatz for 
the ten-dimensional metric, which preserves the five-dimensional
general covariance:
\beqa
ds_{10}^{\:2}&=&G_{MN} \: dX^M dX^N \n
&=& h_{\alpha\beta}(\xi) \: d\xi^{\alpha}d\xi^{\beta}
+e^{\rho(\xi)/2} \: d\Omega_5.
\label{metricansatz}
\eeqa
Here $h_{\alpha\beta}$ is a five-dimensional metric.
We also adopt the following ansatz for the other fields:
\beqa
&&\Phi = \phi (\xi), \n
&&B_2 =\frac{1}{2} \: B_{\alpha\beta}(\xi)\: d\xi^{\alpha} \wedge d\xi^{\beta}
\equiv B, \n
&&C_0 =\chi(\xi), \n
&&C_2= \frac{1}{2} \: C_{\alpha\beta}(\xi)\: d\xi^{\alpha} \wedge d\xi^{\beta} 
\equiv C
\label{matteransatz}
\eeqa
and
\beqa
C_4&=&\frac{1}{4!} \: D_{\alpha \beta \gamma \delta}(\xi) \:
d\xi^{\alpha} \wedge d\xi^{\beta} \wedge d\xi^{\gamma} \wedge d\xi^{\delta}
+\frac{1}{4!} \: k \: E_{\theta_i \theta_j \theta_k \theta_l}(\theta)\:
d\theta_i \wedge d\theta_j \wedge d\theta_k \wedge d\theta_l \n
&\equiv& D + k \: E,
\label{C4}
\eeqa
such that
\beq
5 \: \pa_{[\theta_i}E_{\theta_j \theta_k \theta_l \theta_m]}
=\varepsilon_{\theta_i \theta_j \theta_k \theta_l \theta_m},
\label{derivativeofEtheta}
\eeq
where $\varepsilon_{\theta_i \theta_j \theta_k \theta_l \theta_m}$ is the
totally anti-symmetric covariant tensor in $S^5$ and $k$ is a constant.
We have also defined $B$, $C$, $D$ and $E$: $B$ and $C$ are 2-forms in the 
five dimensions, $D$ is a 4-form in these five dimensions $\xi^{\alpha}$, 
and $E$ is a 4-form
in $S^5$. We set all the other fields to zero. 
We will check below that the ansatz (\ref{C4}) 
is consistent with the equations of motion and the self-duality condition.. From the ansatz (\ref{C4}), 
$\tilde{F}_5$ can be evaluated as
\beq
\tilde{F}_5=\tilde{G} 
+ \frac{1}{5!} \: k \:
\varepsilon_{\theta_i \theta_j \theta_k \theta_l \theta_m} \:
d\theta_i \wedge d\theta_j \wedge d\theta_k \wedge d\theta_l \wedge d\theta_m,
\label{F5tilde}
\eeq
where $\tilde{G}$ is defined by $\tilde{G}=G+C \wedge H$ with
$H=\frac{1}{2} \: \pa_{[\alpha}B_{\beta\gamma]} \:
d\xi^{\alpha} \wedge d\xi^{\beta} \wedge d\xi^{\gamma}$ and
$G=\frac{1}{4!}\: \pa_{[\alpha_1}D_{\alpha_2 \cdots \alpha_5]} \:
d\xi^{\alpha_1} \wedge \cdots \wedge d\xi^{\alpha_5}$. 

We substitute these ansatzes into the equations of motion in
type IIB supergravity (A.1)-(A.6). 
By using the formulae in appendix B, we obtain
the following equations in the five dimensions:
\begin{eqnarray}
&&R^{(5)}_{\alpha \beta}+2\nabla^{(5)}_{\alpha}\nabla^{(5)}_{\beta}\phi
-\frac{5}{4}\nabla^{(5)}_{\alpha}\nabla^{(5)}_{\beta}\rho
-\frac{5}{16} \pa_{\alpha}\rho\pa_{\beta}\rho
-\frac{1}{4}H_{\alpha\gamma\delta}H_{\beta}^{\;\;\gamma\delta} 
-\frac{1}{2}e^{2\phi}\pa_{\alpha}\chi\pa_{\beta}\chi
-\frac{1}{4}e^{2\phi}
\tilde{F}_{\alpha\gamma\delta}\tilde{F}_{\beta}^{\;\;\gamma\delta} \n
&&-\frac{1}{96}e^{2\phi}\tilde{G}_{\alpha\gamma_1 \cdots \gamma_4}
\tilde{G}_{\beta}^{\;\;\gamma_1 \cdots \gamma_4} 
+h_{\alpha \beta} \left(-\frac{1}{2}R^{(5)}
-2\nabla^{(5)}_{\gamma}\nabla^{(5)\gamma}\phi
+\frac{5}{4}\nabla^{(5)}_{\gamma}\nabla^{(5)\gamma}\rho
+2(\pa\phi)^2
+\frac{15}{16}(\pa\rho)^2 \right. \n
&&-\frac{5}{2}\pa_{\gamma}\phi\pa^{\gamma}\rho 
\left. +\frac{1}{4}|H|^2
+\frac{1}{4}e^{2\phi}(\pa\chi)^2
+\frac{1}{4}e^{2\phi}|\tilde{F}|^2
-\frac{1}{2}e^{-\rho/2}R^{(S^5)} \right) =0, \n
&&\mbox{}\n
&&R^{(5)}+4\nabla^{(5)}_\alpha\nabla^{(5)\alpha}\phi
-\frac{5}{2}\nabla^{(5)}_\alpha\nabla^{(5)\alpha}\rho
-4(\pa\phi)^2
-\frac{15}{8}(\pa\rho)^2 
+5\pa_{\alpha}\phi\pa^{\alpha}\rho
-\frac{1}{2}|H|^2 
+e^{-\rho/2}R^{(S^5)} =0,\n
&&\mbox{}\n
&&R^{(5)}+4\nabla^{(5)}_{\alpha}\nabla^{(5)\alpha}\phi
-2\nabla^{(5)}_\alpha\nabla^{(5)\alpha}\rho
-4(\pa\phi)^2 
-\frac{5}{4}(\pa\rho)^2 
+4\pa_{\alpha}\phi\pa^{\alpha}\rho
-\frac{1}{2}|H|^2 \n
&&-\frac{1}{2}e^{2\phi}(\pa\chi)^2
-\frac{1}{2}e^{2\phi}|\tilde{F}|^2
-\frac{1}{2}e^{2\phi}|\tilde{G}|^2
+\frac{3}{5}e^{-\rho/2}R^{(S^5)} =0,\n
&&\mbox{}\n
&&\nabla^{(5)}_{\gamma}(e^{-2\phi+\frac{5}{4}\rho}H^{\gamma\alpha\beta}) 
+\nabla^{(5)}_{\gamma}(e^{\frac{5}{4}\rho}\chi\tilde{F}^{\gamma\alpha\beta}) 
+\frac{1}{6}e^{\frac{5}{4}\rho}F_{\gamma_1 \gamma_2 \gamma_3}
\tilde{G}^{\alpha\beta\gamma_1 \gamma_2 \gamma_3} =0, \n
&&\mbox{}\n
&&\nabla^{(5)}_{\alpha}(e^{\frac{5}{4}\rho}\pa^{\alpha}\chi) 
-\frac{1}{6}e^{\frac{5}{4}\rho}H_{\alpha\beta\gamma} 
\tilde{F}^{\alpha\beta\gamma} = 0, \n
&&\mbox{}\n
&&\nabla^{(5)}_{\gamma}(e^{\frac{5}{4}\rho}\tilde{F}^{\gamma\alpha\beta})
-\frac{1}{6}e^{\frac{5}{4}\rho}H_{\gamma_1 \gamma_2 \gamma_3}
\tilde{G}^{\alpha\beta\gamma_1 \gamma_2 \gamma_3} =0, \n
&&\mbox{}\n
&&\nabla^{(5)}_{\gamma}(e^{\frac{5}{4}\rho}
\tilde{G}^{\gamma\alpha_1 \cdots \alpha_4})=0,
\label{5Deom}
\eeqa
where 
\beqa
&&H=\frac{1}{2} \: \pa_{[\alpha}B_{\beta\gamma]} \:
d\xi^{\alpha} \wedge d\xi^{\beta} \wedge d\xi^{\gamma}, \n
&&F=\frac{1}{2} \: \pa_{[\alpha}C_{\beta\gamma]} \:
d\xi^{\alpha} \wedge d\xi^{\beta} \wedge d\xi^{\gamma}, \n
&&G=\frac{1}{4!}\: \pa_{[\alpha_1}D_{\alpha_2 \cdots \alpha_5]} \:
d\xi^{\alpha_1} \wedge \cdots \wedge d\xi^{\alpha_5}, \n
&&\tilde{F}=F + \chi \wedge H, \n
&&\tilde{G}=G + C \wedge H. 
\eeqa
In the above equations, $|L_q|^2=
h^{\alpha_1\beta_1} \cdots h^{\alpha_q\beta_q}L_{\alpha_1 \cdots \alpha_q}
L_{\beta_1 \cdots \beta_q}$ for a $q$-form $L_q$,
$(\pa\phi)^2=h^{\alpha\beta}\pa_{\alpha}\phi\pa_{\beta}\phi$ and so on.
On the other hand, the self-duality condition (\ref{self-duality}) gives
the relation
\beqa
(\tilde{F}_5)_{\theta_i \theta_j \theta_k \theta_l \theta_m}
&=&\frac{1}{5!}\:\varepsilon_{\theta_i \theta_j \theta_k \theta_l \theta_m}
^{\;\;\;\;\;\;\;\;\;\;\;\;\;\;\;\;\;\;\alpha_1 \cdots \alpha_5} 
\:(\tilde{F}_5)_{\alpha_1 \cdots \alpha_5} \n
&=&-\frac{e^{5\rho/4}}{\sqrt{-h}} \:\tilde{G}_{01234}
\:\varepsilon_{\theta_i \theta_j \theta_k \theta_l \theta_m}.
\label{self-dualityrelation}
\eeqa
By comparing (\ref{F5tilde}) and (\ref{self-dualityrelation}), we obtain
\beq
k=-\frac{e^{5\rho/4}}{\sqrt{-h}} \:\tilde{G}_{01234}.
\label{f}
\eeq
The last equation in (\ref{5Deom}) implies 
that the right-hand side of (\ref{f})
is constant, so that we have verified that the ansatz (\ref{C4}) 
is consistent with the equations of motion and the self-duality condition..

One can easily verify that the equations (\ref{5Deom}) can be derived from
\begin{eqnarray}
&&I_5 = \frac{1}{2\kappa_5^2} \int d^5 \xi \sqrt{-h} 
\left[e^{-2\phi+\frac{5}{4}\rho}
\left( R^{(5)}+4\partial_{\alpha}\phi\partial^{\alpha}\phi 
+\frac{5}{4}\partial_{\alpha}\rho\partial^{\alpha}\rho
-5\partial_{\alpha}\phi\partial^{\alpha}\rho 
-\frac{1}{2}|H|^2 \right) \right.\n
&& \qquad \qquad \qquad \qquad \qquad  \left.
-\frac{1}{2}e^{\frac{5}{4}\rho} \left(
\pa_{\alpha}\chi\pa^{\alpha}\chi
+|\tilde{F}|^2+|\tilde{G}|^2 \right) 
+e^{-2\phi+\frac{3}{4}\rho}R^{(S^5)} \right],
\label{5Daction}
\end{eqnarray}
where 
\beqa
\frac{1}{2\kappa_5^2}=\frac{\mbox{volume of }S^5}{2\kappa_{10}^{\:2}}, \;\;\; 
R^{(S^5)}=20.
\nonumber
\eeqa
Note that by substituting (\ref{metricansatz})
and (\ref{matteransatz}) into the ten-dimensional action (\ref{10Daction}),
one can obtain the above action, except for 
$|\tilde{G}|^2$. We have thus reduced type IIB supergravity on $S^5$ and
obtained the five-dimensional system.
This reduction is a consistent truncation in the sense
that every solution of (\ref{5Daction}) can be lifted to
a solution of type IIB supergravity in ten dimensions. In the remainder of
this paper, we set $2\kappa_5^2=1$.

\vspace{1cm}

\section{ADM formalism and the H-J equation}
\setcounter{equation}{0}
In this section, we develop the ADM formalism for the five-dimensional
system described by (\ref{5Daction}) and derive the H-J
equation. First, we rename the five-dimensional coordinates as follows:
\beqa
\xi^{\mu}=x^{\mu} \;\; (\mu=0,\cdots,3), \;\;\; \xi^4=r.
\nonumber
\eeqa
Adopting $r$ as the time, we carry out
the ADM decomposition for the five-dimensional
metric
\beqa
ds_5^2&=&h_{\alpha\beta} \: d\xi^{\alpha}d\xi^{\beta} \n
&=&(n^2+g^{\mu\nu}n_{\mu}n_{\nu})\:dr^2+2n_{\mu}\:dr\:dx^{\mu}
+g_{\mu\nu}\:dx^{\mu}dx^{\nu},
\label{ADMdecomposition}
\eeqa
where $n$ and $n_{\mu}$ are the lapse function and the shift function,
respectively. Henceforth $\mu$ and $\nu$ run from 0 to 3.

In what follows, we consider a boundary surface 
specified by $r=\mbox{const.}$ and impose 
the Dirichlet condition for the fields on the boundary. Here we need to
add the Gibbons-Hawking term \cite{GH} to (\ref{5Daction}), 
which is defined on the
boundary and ensures that the Dirichlet condition can be 
imposed consistently \cite{FMS,NOO,MM}.
Then, the five-dimensional action (\ref{5Daction}) with 
the Gibbons-Hawking term on the boundary can be expressed 
in terms of the ADM variables as
\beqa
&&I_5=\int dr d^4x \sqrt{-g} n \left[
e^{-2\phi+\frac{5}{4}\rho} \left(
-(K_{\mu\nu})^2+K^2 \right.\right.\n
&&\qquad \qquad \qquad \qquad \qquad  
+\frac{1}{n}\left(-4(\pa_r \phi-n^{\mu}\pa_{\mu}\phi)
+\frac{5}{2}(\pa_r \rho-n^{\mu}\pa_{\mu}\rho)\right)K  \n
&&\qquad \qquad \qquad \qquad \qquad  
+\frac{1}{n^2}\left(4(\pa_r \phi-n^{\mu}\pa_{\mu}\phi)^2
+\frac{5}{4}(\pa_r \rho-n^{\mu}\pa_{\mu}\rho)^2 \right.\n
&&\qquad \qquad \qquad \qquad \qquad  \left.\left.
-5(\pa_r \phi-n^{\mu}\pa_{\mu}\phi)(\pa_r \rho-n^{\mu}\pa_{\mu}\rho)
-\frac{1}{4}(H_{r\mu\nu}-n^{\lambda}H_{\lambda\mu\nu})^2 \right)\right) \n
&&\qquad \qquad \qquad \qquad \qquad  
+\frac{1}{n^2}e^{\frac{5}{4}\rho}\left( 
-\frac{1}{2}(\pa_r \chi-n^{\mu}\pa_{\mu}\chi)^2
-\frac{1}{4}(\tilde{F}_{r\mu\nu}-n^{\lambda}\tilde{F}_{\lambda\mu\nu})^2 
\right.\n
&&\qquad \qquad \qquad \qquad \qquad  \left.\left. 
-\frac{1}{48}(\tilde{G}_{r\mu\nu\lambda\rho}
-n^{\sigma}\tilde{G}_{\sigma\mu\nu\lambda\rho})^2 \right) +\cal{L} \;\;\right],
\label{I5}
\eeqa
where
\beqa
&&{\cal L}=e^{-2\phi+\frac{5}{4}\rho} \left(R_g+4\nabla_{\mu}\nabla^{\mu}\phi
-\frac{5}{2}\nabla_{\mu}\nabla^{\mu}\rho-4\pa_{\mu}\phi\pa^{\mu}\phi
-\frac{15}{8}\pa_{\mu}\rho\pa^{\mu}\rho
+5\pa_{\mu}\phi\pa^{\mu}\rho
-\frac{1}{12}H_{\mu\nu\lambda}H^{\mu\nu\lambda} \right) \n
&& \qquad +e^{\frac{5}{4}\rho} \left( -\frac{1}{2}\pa_{\mu}\chi\pa^{\mu}\chi
-\frac{1}{12}\tilde{F}_{\mu\nu\lambda}\tilde{F}^{\mu\nu\lambda} \right) 
+e^{-2\phi+\frac{3}{4}\rho}R^{(S^5)},
\label{L}
\eeqa
and $K_{\mu\nu}$ is the extrinsic curvature on the four-dimensional
manifold given by
\beq
K_{\mu\nu}=\frac{1}{2n}(\pa_r g_{\mu\nu}-\nabla_{\mu}n_{\nu}
-\nabla_{\nu}n_{\mu}), \;\;\;
K=g^{\mu\nu}K_{\mu\nu}.
\eeq

Furthermore, by introducing the canonical momenta, we rewrite the 
above expression as
\beqa
&&I_5 = \int dr d^4x 
\sqrt{-g}(\pi^{\mu\nu}\pa_r g_{\mu\nu}+\pi_{\phi}\pa_r \phi
+\pi_{\rho}\pa_r \rho+\pi_B^{\mu\nu} \pa_r B_{\mu\nu} \n
&&\qquad \qquad \qquad \qquad   \;\;+\pi_{\chi}\pa_r \chi
+\pi_C^{\mu\nu} \pa_r C_{\mu\nu} 
+\pi_D^{\mu\nu\lambda\rho} \pa_r D_{\mu\nu\lambda\rho}\n
&&\qquad \qquad \qquad \qquad   \;\;
-nH-n_{\mu}H^{\mu}-B_{r\mu}Z_B^{\mu}-C_{r\mu}Z_C^{\mu}
-D_{r\mu\nu\lambda}Z_D^{\mu\nu\lambda}),
\label{ADM}
\eeqa
with
\beqa
&&H=-e^{2\phi-\frac{5}{4}\rho} \left( (\pi^{\mu\nu})^2
+\frac{1}{2}{\pi_{\phi}}^2+\frac{1}{2}\pi^{\mu}_{\;\; \mu}\pi_{\phi}
+\frac{4}{5}{\pi_{\rho}}^2+\pi_{\phi}\pi_{\rho} 
+\left( \pi_B^{\mu\nu}-\chi\pi_C^{\mu\nu}
-6C_{\lambda\rho}\pi_D^{\mu\nu\lambda\rho} \right)^2 \right) \n
&& \qquad -e^{-\frac{5}{4}\rho} \left( \frac{1}{2}{\pi_{\chi}}^2
+(\pi_C^{\mu\nu})^2 +12(\pi_D^{\mu\nu\lambda\rho})^2  \right)
- {\cal L}, \\
&&H^{\mu}=-2\nabla_{\nu}\pi^{\mu\nu}+\pi_{\phi}\pa^{\mu}\phi
+\pi_{\rho}\pa^{\mu}\rho +\pi_{B\nu\lambda} H^{\mu\nu\lambda} \n
&& \qquad \;\; +\pi_{\chi}\pa^{\mu}\chi
+\pi_{C\nu\lambda}F^{\mu\nu\lambda}
+\pi_{D\nu\lambda\rho\sigma}(G^{\mu\nu\lambda\rho\sigma}
+4C^{\mu\nu}H^{\lambda\rho\sigma}),\\
&&Z_B^{\mu}=2\nabla_{\nu}\pi_B^{\mu\nu}, \\
&&Z_C^{\mu}=2\nabla_{\nu}\pi_C^{\mu\nu}
-4\pi_D^{\mu\nu\lambda\rho}H_{\nu\lambda\rho}, \\
&&Z_D^{\mu\nu\lambda}=4\nabla_{\rho}\pi^{\mu\nu\lambda\rho}.
\eeqa
In fact, by varying (\ref{ADM}) with respect to $\pi^{\mu\nu},\;
\pi_{\phi}, \; \pi_{\rho}, \; \pi_B^{\mu\nu}, \pi_{\chi}$,
$\pi_C^{\mu\nu}$ and $\pi_D^{\mu\nu\lambda\rho}$, we obtain the relations
\beqa
&&\pi_{\mu\nu}=e^{-2\phi+\frac{5}{4}\rho} \left( -K_{\mu\nu}+g_{\mu\nu}K
-\frac{2}{n}g_{\mu\nu}(\pa_r \phi-n^{\lambda}\pa_{\lambda}\phi)
+\frac{5}{4n}g_{\mu\nu}(\pa_r \rho-n^{\lambda}\pa_{\lambda}\rho)  \right),\n
&&\pi_{\phi}=e^{-2\phi+\frac{5}{4}\rho} 
\left( -4K+\frac{8}{n}(\pa_r \phi-n^{\mu}\pa_{\mu}\phi)
-\frac{5}{n}(\pa_r \rho-n^{\mu}\pa_{\mu}\rho) \right), \n
&&\pi_{\rho}=e^{-2\phi+\frac{5}{4}\rho} 
\left( \frac{5}{2}K-\frac{5}{n}(\pa_r \phi-n^{\mu}\pa_{\mu}\phi)
+\frac{5}{2n}(\pa_r \rho-n^{\mu}\pa_{\mu}\rho) \right), \n
&&\pi_{B\mu\nu}=\frac{1}{n}\left(
-\frac{1}{2}e^{-2\phi+\frac{5}{4}\rho}
(H_{r\mu\nu}-n^{\lambda}H_{\lambda\mu\nu})
-\frac{1}{2}e^{\frac{5}{4}\rho}
\chi(\tilde{F}_{r\mu\nu}-n^{\lambda}\tilde{F}_{\lambda\mu\nu}) \right.\n
&&\qquad \qquad \left. -\frac{1}{4}e^{\frac{5}{4}\rho}C^{\lambda\rho}
(\tilde{G}_{r\mu\nu\lambda\rho}-n^{\sigma}\tilde{G}_{\sigma\mu\nu\lambda\rho})
\right), \n
&&\pi_{\chi}=-\frac{1}{n}e^{\frac{5}{4}\rho}
(\pa_r \chi-n^{\mu}\pa_{\mu}\chi), \n
&&\pi_{C\mu\nu}=-\frac{1}{2n}e^{\frac{5}{4}\rho}
(\tilde{F}_{r\mu\nu}-n^{\lambda}\tilde{F}_{\lambda\mu\nu}), \n
&&\pi_{D\mu\nu\lambda\rho}=-\frac{1}{24n}e^{\frac{5}{4}\rho}
(\tilde{G}_{r\mu\nu\lambda\rho}-n^{\sigma}\tilde{G}_{\sigma\mu\nu\lambda\rho}),
\label{piandfield}
\eeqa 
and by substituting these relations into (\ref{ADM}), we reproduce
(\ref{I5}). 

Here $n,\; n_{\mu}, \; B_{r\mu}$, $C_{r\mu}$ and $D_{r\mu\nu\lambda}$ behave
like Lagrange multipliers, giving the constraints
\beq
H=0, \;\;\; H^{\mu}=0, \;\;\; Z_B^{\mu}=0, \;\;\;
Z_C^{\mu}=0, \;\;\; Z_D^{\mu\nu\lambda}=0.
\label{constraints}
\eeq
The first of these is the Hamiltonian constraint, the second is the
momentum constraint and the last three are the Gauss law constraints
coming from the $U(1)$ gauge symmetries for $B$, $C$ and $D$.

In the remainder of this section, we derive the H-J equation.
Let $\bar{g}_{\mu\nu}(x,r)$, 
$\bar{\phi}(x,r)$, $\bar{\rho}(x,r)$, 
$\bar{B}_{\mu\nu}(x,r)$,
$\bar{\chi}(x,r)$, $\bar{C}_{\mu\nu}(x,r)$ and
$\bar{D}_{\mu\nu\lambda\rho}(x,r)$
be a classical solution of (\ref{I5}) with the boundary conditions
\beqa
&&\bar{g}_{\mu\nu}(x,r=r_0)=g_{\mu\nu}(x),\;\;\;
\bar{\phi}(x,r=r_0)=\phi(x),\;\;\;
\bar{\rho}(x,r=r_0)=\rho(x), \n
&&\bar{B}_{\mu\nu}(x,r=r_0)=B_{\mu\nu}(x),\;\;\;
\bar{\chi}(x,r=r_0)=\chi(x),\;\;\;
\bar{C}_{\mu\nu}(x,r=r_0)=C_{\mu\nu}(x),\n
&&\bar{D}_{\mu\nu\lambda\rho}(x,r=r_0)=D_{\mu\nu\lambda\rho}(x).
\label{bc}
\eeqa
We also define $\pi_{\mu\nu}(x),\cdots,\pi_D^{\mu\nu\lambda\rho}(x)$
by
\beqa
&&\pi^{\mu\nu}(x)=\bar{\pi}^{\mu\nu}(x,r=r_0),\;\;\;
\pi_{\phi}(x)=\bar{\pi}_{\phi}(x,r=r_0),\;\;\;
\pi_{\rho}(x)=\bar{\pi}_{\rho}(x,r=r_0),\n
&&\pi_B^{\mu\nu}(x)=\bar{\pi}_B^{\mu\nu}(x,r=r_0),\;\;\;
\pi_{\chi}(x)=\bar{\pi}_{\chi}(x,r=r_0),\;\;\;
\pi_C^{\mu\nu}(x)=\bar{\pi}_C^{\mu\nu}(x,r=r_0), \n
&&\pi_D^{\mu\nu\lambda\rho}(x)=\bar{\pi}_D^{\mu\nu\lambda\rho}(x,r=r_0),
\label{pibar}
\eeqa
where 
the right-hand sides of these equations are calculated using the relations
(\ref{piandfield}) for the classical solution.

By substituting the solution into (\ref{I5})
with the boundary specified by $r=r_0$, 
we obtain the on-shell action $S$, 
which is in general a functional
of $g_{\mu\nu}(x),\cdots,D_{\mu\nu\lambda\rho}(x)$ and $r_0$.
The standard argument employed in the H-J formalism leads to 
the relations (for example, see Ref.\cite{FMS})
\beqa
&&\pi^{\mu\nu}(x)=\frac{1}{\sqrt{-g(x)}}
\frac{\delta S}{\delta g_{\mu\nu}(x)},\;\;\;
\pi_{\phi}(x)=\frac{1}{\sqrt{-g(x)}}
\frac{\delta S}{\delta \phi(x)},\;\;\;
\pi_{\rho}(x)=\frac{1}{\sqrt{-g(x)}}
\frac{\delta S}{\delta \rho(x)},\n
&&\pi_{B}^{\mu\nu}(x)=\frac{1}{\sqrt{-g(x)}}
\frac{\delta S}{\delta B_{\mu\nu}(x)},\;\;\;
\pi_{\chi}(x)=\frac{1}{\sqrt{-g(x)}}
\frac{\delta S}{\delta \chi(x)},\;\;\;
\pi_{C}^{\mu\nu}(x)=\frac{1}{\sqrt{-g(x)}}
\frac{\delta S}{\delta C_{\mu\nu}(x)}, \n
&&\pi_{D}^{\mu\nu\lambda\rho}(x)=\frac{1}{\sqrt{-g(x)}}
\frac{\delta S}{\delta D_{\mu\nu\lambda\rho}(x)}
\label{pianddelS}
\eeqa
and
\beq
\frac{\pa S}{\pa r_0}=0.
\eeq
The last equation is characteristic of gravitational systems; that is,
the on-shell action does not depend on the boundary time explicitly.

The quantities 
$\bar{g}_{\mu\nu}(x,r=r_0),\cdots,\bar{D}_{\mu\nu\lambda\rho}(x,r=r_0)$ and
$\bar{\pi}^{\mu\nu}(x,r=r_0),\cdots,\bar{\pi}_D^{\mu\nu\lambda\rho}(x,r=r_0)$
must satisfy the constraints (\ref{constraints}). Therefore we see from
(\ref{bc}), (\ref{pibar}) and (\ref{pianddelS}) 
that the constraints give functional
differential equations for $S$. The momentum constraint and the Gauss law
constraints imply that $S$ must be invariant under the
diffeomorphism in four dimensions and the $U(1)$ gauge transformations.
We give a proof of this in appendix C. On the other hand, 
the Hamiltonian constraint
gives a non-trivial equation that determines the form of $S$. We call
this equation the H-J equation and solve it in the
next section.

\vspace{1cm}

\section{D3-brane effective action as a solution to the H-J 
equation}
\setcounter{equation}{0}

We assume that the fields are constant on the fixed-time
surface. Let $S_0$ be a solution to the H-J equation
under this assumption. Then, we see from (\ref{L}), (3.6) 
and (\ref{pianddelS}) that
$S_0$ satisfies the equation
\beqa
&&-e^{2\phi-\frac{5}{4}\rho} \left(
\left(\frac{1}{\sqrt{-g}}\frac{\delta S_0}{\delta g_{\mu\nu}}\right)^2
+\frac{1}{2}g_{\mu\nu}\frac{1}{\sqrt{-g}}\frac{\delta S_0}{\delta g_{\mu\nu}}
\frac{1}{\sqrt{-g}}\frac{\delta S_0}{\delta \phi}
+\frac{1}{2}\left(\frac{1}{\sqrt{-g}}\frac{\delta S_0}{\delta \phi}\right)^2
+\frac{4}{5}\left(\frac{1}{\sqrt{-g}}\frac{\delta S_0}{\delta \rho}\right)^2 
\right.\n
&&\left. \qquad \qquad +\frac{1}{\sqrt{-g}}\frac{\delta S_0}{\delta \phi}
\frac{1}{\sqrt{-g}}\frac{\delta S_0}{\delta \rho}
+\left(\frac{1}{\sqrt{-g}}\frac{\delta S_0}{\delta B_{\mu\nu}}
-\chi\frac{1}{\sqrt{-g}}\frac{\delta S_0}{\delta C_{\mu\nu}}
-6C_{\lambda\rho}\frac{1}{\sqrt{-g}}
\frac{\delta S_0}{\delta D_{\mu\nu\lambda\rho}}\right)^2 \right) \n
&&-e^{-\frac{5}{4}\rho} \left(
\frac{1}{2}\left(\frac{1}{\sqrt{-g}}\frac{\delta S_0}{\delta \chi}\right)^2
+\left(\frac{1}{\sqrt{-g}}\frac{\delta S_0}{\delta C_{\mu\nu}}\right)^2
+12\left(\frac{1}{\sqrt{-g}}\frac{\delta S_0}{\delta D_{\mu\nu\lambda\rho}}
\right)^2 \right) \n
&&=e^{-2\phi+\frac{3}{4}\rho}R^{(S^5)}.
\label{HJ}
\eeqa

We show that the form
\beq
S_0=S_c+S_{BI}+S_{WZ}+\sigma
\label{S0}
\eeq
is a solution to (\ref{HJ}), with
\beqa
S_c&=&\alpha \int d^4x \sqrt{-g} e^{-2\phi+\rho}, \n
S_{BI}&=&\beta \int d^4x e^{-\phi} \sqrt{-\det (g_{\mu\nu}+{\cal F}_{\mu\nu})}, \n
S_{WZ}&=&\gamma \left( \int D + \int C \wedge {\cal F} 
+ \frac{1}{2}\int \chi \; {\cal F} \wedge {\cal F} \right) \n
&=& \gamma \int d^4x \sqrt{-g} \varepsilon^{\mu\nu\lambda\rho}
\left( \frac{1}{24}D_{\mu\nu\lambda\rho}
+\frac{1}{4}C_{\mu\nu}{\cal F}_{\lambda\rho}
+\frac{1}{8}\chi {\cal F}_{\mu\nu}{\cal F}_{\lambda\rho} \right),
\label{ScSBISWZ}
\eeqa 
where ${\cal F}_{\mu\nu}=B_{\mu\nu}+F_{\mu\nu}$,
$F_{\mu\nu}$ is an arbitrary constant anti-symmetric tensor, and $\sigma$ 
is an  arbitrary constant.
Noting that
\beqa
\frac{1}{\sqrt{-g}}\frac{\delta S_0}{\delta B_{\mu\nu}}
-\chi\frac{1}{\sqrt{-g}}\frac{\delta S_0}{\delta C_{\mu\nu}}
-6C_{\lambda\rho}\frac{1}{\sqrt{-g}}
\frac{\delta S_0}{\delta D_{\mu\nu\lambda\rho}}
=\frac{1}{\sqrt{-g}}\frac{\delta S_{BI}}{\delta B_{\mu\nu}} 
\nonumber
\eeqa
and 
\beqa
\frac{1}{\sqrt{-g}}\frac{\delta S_{WZ}}{\delta g_{\mu\nu}}=0,
\nonumber
\eeqa
one can see that the left-hand side of (\ref{HJ}) can be decomposed
into the four parts
\beq
\mbox{L.H.S. of (\ref{HJ})}=-e^{2\phi-\frac{5}{4}\rho} (\;(1)+(2)+(3)\;)
-e^{-\frac{5}{4}\rho}\times (4),
\label{LHS}
\eeq
with
\beqa
&&(1)=\left(\frac{1}{\sqrt{-g}}\frac{\delta S_c}{\delta g_{\mu\nu}}\right)^2
+\frac{1}{2}g_{\mu\nu}\frac{1}{\sqrt{-g}}\frac{\delta S_c}{\delta g_{\mu\nu}}
\frac{1}{\sqrt{-g}}\frac{\delta S_c}{\delta \phi}
+\frac{1}{2}\left(\frac{1}{\sqrt{-g}}\frac{\delta S_c}{\delta \phi}\right)^2
+\frac{4}{5}\left(\frac{1}{\sqrt{-g}}\frac{\delta S_c}{\delta \rho}\right)^2 \n
&&\qquad \;\; +\frac{1}{\sqrt{-g}}\frac{\delta S_c}{\delta \phi}
\frac{1}{\sqrt{-g}}\frac{\delta S_c}{\delta \rho}, \n
&&(2)=2g_{\mu\lambda}g_{\nu\rho}
\frac{1}{\sqrt{-g}}\frac{\delta S_c}{\delta g_{\mu\nu}}
\frac{1}{\sqrt{-g}}\frac{\delta S_{BI}}{\delta g_{\lambda\rho}}
+\frac{1}{2}g_{\mu\nu}\frac{1}{\sqrt{-g}}\frac{\delta S_c}{\delta g_{\mu\nu}}
\frac{1}{\sqrt{-g}}\frac{\delta S_{BI}}{\delta \phi} \n
&&\qquad \;\; +\frac{1}{2}g_{\mu\nu}
\frac{1}{\sqrt{-g}}\frac{\delta S_{BI}}{\delta g_{\mu\nu}}
\frac{1}{\sqrt{-g}}\frac{\delta S_c}{\delta \phi}
+\frac{1}{\sqrt{-g}}\frac{\delta S_c}{\delta \phi}
\frac{1}{\sqrt{-g}}\frac{\delta S_{BI}}{\delta \phi} 
+\frac{1}{\sqrt{-g}}\frac{\delta S_{BI}}{\delta \phi}
\frac{1}{\sqrt{-g}}\frac{\delta S_c}{\delta \rho}, \n
&&(3)=\left(\frac{1}{\sqrt{-g}}\frac{\delta S_{BI}}{\delta g_{\mu\nu}}\right)^2
+\frac{1}{2}g_{\mu\nu}\frac{1}
{\sqrt{-g}}\frac{\delta S_{BI}}{\delta g_{\mu\nu}}
\frac{1}{\sqrt{-g}}\frac{\delta S_{BI}}{\delta \phi}
+\frac{1}{2}\left(\frac{1}{\sqrt{-g}}
\frac{\delta S_{BI}}{\delta \phi}\right)^2 
+\left(\frac{1}{\sqrt{-g}}\frac{\delta S_{BI}}{\delta B_{\mu\nu}}\right)^2, \n
&&(4)=
\frac{1}{2}\left(\frac{1}{\sqrt{-g}}\frac{\delta S_{WZ}}{\delta \chi}\right)^2
+\left(\frac{1}{\sqrt{-g}}\frac{\delta S_{WZ}}{\delta C_{\mu\nu}}\right)^2
+12\left(\frac{1}{\sqrt{-g}}
\frac{\delta S_{WZ}}{\delta D_{\mu\nu\lambda\rho}}\right)^2.
\eeqa
(1) and (4) are easily calculated as
\beqa
(1)&=&-\frac{1}{5}\alpha^2 e^{-4\phi+2\rho}, \n
(4)&=&-\frac{1}{2}\gamma^2 \left( 1
+\frac{1}{2}{\cal F}_{\mu\nu}{\cal F}^{\mu\nu}
+\frac{1}{8}({\cal F}_{\mu\nu}{\cal F}^{\mu\nu})^2
-\frac{1}{4}{\cal F}_{\mu\nu}{\cal F}^{\nu\lambda}
{\cal F}_{\lambda\rho}{\cal F}^{\rho\mu} \right).
\label{(1)and(4)}
\eeqa

In order to calculate (2) and (3), we introduce the $4 \times 4$
matrices ${\cal G}$
and ${\cal B}$:
\beqa
({\cal G})_{\mu\nu}=g_{\mu\nu}, \;\;\;
({\cal B})_{\mu\nu}={\cal F}_{\mu\nu}.
\nonumber
\eeqa
Then, we have
\beqa
&&\frac{1}{\sqrt{-g}}\frac{\delta S_c}{\delta g_{\mu\nu}}
=\frac{1}{2}\alpha e^{-2\phi+\rho} \left( \frac{1}{{\cal G}} \right)^{\mu\nu},
\;\;\;
\frac{1}{\sqrt{-g}}\frac{\delta S_c}{\delta \phi}
=-2\alpha e^{-2\phi+\rho}, 
\;\;\;
\frac{1}{\sqrt{-g}}\frac{\delta S_c}{\delta \rho}
=\alpha e^{-2\phi+\rho}, \n
&&\frac{1}{\sqrt{-g}}\frac{\delta S_{BI}}{\delta g_{\mu\nu}}
=\frac{1}{2}\beta e^{-\phi} 
\sqrt{\frac{\det ({\cal G}+{\cal B})}{\det {\cal G}}}
\left( \frac{1}{{\cal G}+{\cal B}} \: {\cal G} \: \frac{1}{{\cal G}-{\cal B}}
\right)^{\mu\nu}, \n
&&\frac{1}{\sqrt{-g}}\frac{\delta S_{BI}}{\delta B_{\mu\nu}}
=\frac{1}{2}\beta e^{-\phi} 
\sqrt{\frac{\det ({\cal G}+{\cal B})}{\det {\cal G}}}
\left( \frac{1}{{\cal G}+{\cal B}} \: {\cal B} \: \frac{1}{{\cal G}-{\cal B}}
\right)^{\mu\nu}, \n
&&\frac{1}{\sqrt{-g}}\frac{\delta S_{BI}}{\delta \phi}
=-\beta e^{-\phi} 
\sqrt{\frac{\det ({\cal G}+{\cal B})}{\det {\cal G}}}.
\eeqa
Using this notation, we can express each term in (2) and (3) in terms
of the trace of the $4 \times 4$ matrix and calculate (2) and (3) as follows: 
\beqa
(2)&=&\alpha\beta e^{-3\phi+\rho} 
\sqrt{\frac{\det ({\cal G}+{\cal B})}{\det {\cal G}}} \left( 
\frac{1}{2}\tr \left(\frac{1}{{\cal G}} \: {\cal G} \: 
\frac{1}{{\cal G}+{\cal B}}
\: {\cal G} \: \frac{1}{{\cal G}-{\cal B}} \: {\cal G} \right) -1 \right. \n
&&\qquad\qquad\qquad\qquad\qquad\;\;\; \left.
-\frac{1}{2}\tr \left(\frac{1}{{\cal G}+{\cal B}}
\: {\cal G} \: \frac{1}{{\cal G}-{\cal B}} \: {\cal G} \right) +2-1 \right) \n 
&=&0, \n
(3)&=&\frac{1}{4}\beta^2 e^{-2\phi} 
\frac{\det ({\cal G}+{\cal B})}{\det {\cal G}} \left(
\tr \left(\frac{1}{{\cal G}+{\cal B}} \: {\cal G} \: 
\frac{1}{{\cal G}-{\cal B}} \: {\cal G} \:
\frac{1}{{\cal G}+{\cal B}} \: {\cal G} \:
\frac{1}{{\cal G}-{\cal B}} \: {\cal G} \right) 
-\tr \left(\frac{1}{{\cal G}+{\cal B}} \: {\cal G} \: 
\frac{1}{{\cal G}-{\cal B}} \: {\cal G} \right) \right.\n
&&\qquad\qquad\qquad\qquad\;\;\;\; \left.
+2 
-\tr \left(\frac{1}{{\cal G}+{\cal B}} \: {\cal B} \: 
\frac{1}{{\cal G}-{\cal B}} \: {\cal G} \:
\frac{1}{{\cal G}+{\cal B}} \: {\cal B} \:
\frac{1}{{\cal G}-{\cal B}} \: {\cal G} \right) \right) \n
&=&\frac{1}{2}\beta^2 e^{-2\phi} 
\frac{\det ({\cal G}+{\cal B})}{\det {\cal G}} \n
&=&\frac{1}{2}\beta^2 e^{-2\phi} 
\left( 1+\frac{1}{2}{\cal F}_{\mu\nu}{\cal F}^{\mu\nu}
+\frac{1}{8}({\cal F}_{\mu\nu}{\cal F}^{\mu\nu})^2
-\frac{1}{4}{\cal F}_{\mu\nu}{\cal F}^{\nu\lambda}
{\cal F}_{\lambda\rho}{\cal F}^{\rho\mu} \right).
\label{(2)and(3)}
\eeqa
$\mbox{}$ From (\ref{HJ}), (\ref{LHS}), (\ref{(1)and(4)}) 
and (\ref{(2)and(3)}), we conclude that $S_0$
satisfies the H-J equation (\ref{HJ}) if 
\beq
\alpha^2=5 \: R^{(S^5)}=100 \;\;\; \mbox{and} \;\;\;  \beta^2=\gamma^2.
\label{condition}
\eeq

\vspace*{1cm}

\section{Supergravity solution of D3-branes in $B_2$ field}
\setcounter{equation}{0}
In this section, we first review  
the supergravity solution of $N$ D3-branes 
in a constant $B_2$ field background \cite{RT-BMM} and its near-horizon limit \cite{HI-MR}.  
Next, we find that
the on-shell actions for this supergravity solution and 
its near-horizon limit are included in our solution to the H-J equation
obtained in the previous section.

The supergravity solution of $N$ D3-branes 
with only $(B_2)_{23}$ non-vanishing that
preserves 16 supersymmetries \cite{RT-BMM} is also a solution of the
five-dimensional gravity (\ref{I5}) given by
\beqa
&&ds_{5}^{2}=f^{-1/2} [-(dx^0)^2+(dx^1)^2+h((dx^2)^2+(dx^3)^2)]
+f^{1/2}dr^2, \n
&& f=1+\frac{{\alpha'}^2 R^4}{r^4}, \;\;\; 
h^{-1}=\sin^2 \theta f^{-1} + \cos^2 \theta, \n
&& e^{2\phi}=g^2 h, \;\;\; e^{\rho/2}=r^2f^{1/2}, \;\;\;
B_{23}=\tan \theta f^{-1} h, \n
&& C_{01}=\frac{1}{g}\sin\theta f^{-1}, \;\;\;\; 
D_{0123}=\frac{\cos\theta}{g} f^{-1} h, \n
&&\cos\theta R^4=4\pi g N.
\label{D3branesolution}
\eeqa
The ten-dimensional geometry of this solution 
is asymptotic to flat space for $r \rightarrow \infty$,
while it has a horizon
at $r=0$ and behaves like $AdS_5 \times S^5$ near $r=0$.
When $\theta=0$, the solution reduces to the ordinary D3-brane solution.

In order to decouple the asymptotic region with the $B$ field
remaining non-trivial,
the parameters should be rescaled as
\beqa
&&\alpha' \rightarrow 0, \;\;\; \tan\theta=\frac{\tilde{b}}{\alpha'}, \n
&&x^{0,1} \rightarrow x^{0,1}, \;\;\; 
\frac{\tilde{b}}{\alpha'}x^{2,3} \rightarrow x^{2,3}, \n
&&r=\alpha' R^2 u, \;\;\; g=\frac{\alpha'}{\tilde{b}} \hat{g},
\label{scaling}
\eeqa
where $\tilde{b}$, $u$, $\hat{g}$ and the new coordinates $x^{\mu}$ are fixed. 
This scaling corresponds to the Seiberg-Witten limit \cite{SW}. 
Then, the supergravity solution
(\ref{D3branesolution}) reduces to the following form \cite{HI-MR}:
\begin{eqnarray}
&&ds_{5}^{2} = \alpha'R^2 \left[ u^2 (-(dx^0)^2+(dx^1)^2) 
+ \frac{u^2}{1+a^4u^4}((dx^2)^2 + (dx^3)^2) 
+ \frac{du^2}{u^2} \right], \n
&& e^{2\phi}= \hat{g}^2 \frac{1}{1+a^4u^4}, \;\;\;
e^{\rho/2}=\alpha'R^2, \;\;\;
B_{23}=\frac{\alpha'}{\tilde{b}} \frac{a^4u^4}{1+a^4u^4},\n
&& C_{01}=\frac{\alpha'}{\hat{g}\tilde{b}}a^4u^4, \;\;\;
D_{0123}
=\frac{{\alpha'}^2R^4}{\hat{g}}\frac{u^4}{1+a^4u^4}, \n
&&a^2=\tilde{b}R^2, \;\;\; R^4=4\pi \hat{g} N.
\label{background}
\end{eqnarray}
Note that (\ref{background}) is still a solution of the five-dimensional
gravity (\ref{I5}) with the identification $u=r$. 
This solution is conjectured to be
the gravity dual of noncommutative Yang Mills in which $\theta_{23}=\tilde{b}$.
When $\tilde{b}=0$, the ten-dimensional geometry of this solution
is identical to $AdS_5 \times S^5$, which
is dual to the ordinary ${\cal N}=4$ super Yang Mills. 

Let us show that the on-shell actions
for (\ref{D3branesolution}) and (\ref{background}) are reproduced by
the solution (\ref{S0}) to the H-J equation. 
First, by using (\ref{piandfield}),
we calculate the values of the canonical momenta
for (\ref{D3branesolution}) and (\ref{background}) on the boundaries
specified by $r=r_0$ and $u=u_0$, respectively. 
For (\ref{D3branesolution}), we obtain
\beqa
&&-\pi_{00}=\pi_{11}=\frac{f_0^{-1/2}}{g^2 h_0}
\left( 5r_0^4 f_0 +\frac{1}{2}r_0^5\pa_{r_0} f_0 \right), \n
&&\pi_{22}=\pi_{33}=\frac{f_0^{-1/2}}{g^2} 
\left( 5r_0^4 f_0 +\frac{1}{2}r_0^5\pa_{r_0} f_0
-\frac{1}{2}\sin^2 \theta r_0^5\pa_{r_0} f_0 f_0^{-1} h_0 \right), \n
&&\pi_{\phi}=\frac{1}{g^2 h_0}(-20r_0^4 f_0-r_0^5 \pa_{r_0} f_0), \;\;\;
\pi_{\rho}=\frac{10}{g^2 h_0} r_0^4 f_0, \;\;\; \pi_{B23}=0, \n
&&\pi_{\chi}=0, \;\;\; 
\pi_{C01}=\frac{\sin\theta}{2g}r_0^5 f_0^{-1}\pa_{r_0} f_0, \;\;\;
\pi_{D0123}=\frac{\cos\theta}{24g}r_0^5 h_0 f_0^{-1} \pa_{r_0} f_0,
\label{onshellpi1}
\eeqa
where
\beqa
f_0=1+\frac{{\alpha'}^2R^4}{r_0^4}, 
\;\;\; h_0^{-1}=\sin^2 \theta f_0^{-1} +\cos^2 \theta.
\nonumber
\eeqa
For (\ref{background}), we obtain
\beqa
&&-\pi_{00}=\pi_{11}=\frac{3{\alpha'}^3R^6}{\hat{g}^2}
u_0^2 (1+a^4u_0^4), \;\;\;
\pi_{22}=\pi_{33}=\frac{{\alpha'}^3R^6}{\hat{g}^2}
u_0^2 \left( 3+\frac{2a^4u_0^4}{1+a^4u_0^4} \right), \n
&&\pi_{\phi}=-\frac{16{\alpha'}^2R^4}{\hat{g}^2}(1+a^4u_0^4), \;\;\;
\pi_{\rho}=\frac{10{\alpha'}^2R^4}{\hat{g}^2}(1+a^4u_0^4), \;\;\;
\pi_{B23}=0, \n
&&\pi_{\chi}=0, \;\;\;
\pi_{C01}=-2{\alpha'}^2R^4\frac{\alpha'}{\hat{g}\tilde{b}}a^4u_0^4, \;\;\;
\pi_{D0123}=-\frac{{\alpha'}^4R^8}{6\hat{g}}\frac{u_0^4}{1+a^4u_0^4}.
\label{onshellpi2}
\eeqa
Note that the right-hand sides of (\ref{onshellpi1}) reduce to the
right-hand sides of (\ref{onshellpi2}) in the near-horizon 
limit (\ref{scaling}).
On the other hand, the solution (\ref{S0}) to the H-J equation gives 
the following canonical momenta:
\beqa
&&\pi_{\mu\nu}=g_{\mu\lambda}g_{\nu\rho}
\frac{1}{\sqrt{-g}}\frac{\delta S_0}{\delta g_{\lambda\rho}} \n
&&\qquad \!
=\frac{1}{2}(\alpha e^{-2\phi+\rho} +\beta e^{-\phi} A)g_{\mu\nu}
+\frac{\beta e^{-\phi}}{2A}
\left( -{\cal F}_{\mu\lambda}{\cal F}_{\nu}^{\;\;\lambda} 
-\frac{1}{2}{\cal F}_{\mu\lambda}{\cal F}_{\nu}^{\;\;\lambda}
{\cal F}_{\rho\sigma}{\cal F}^{\rho\sigma}
+{\cal F}_{\mu\lambda}{\cal F}^{\lambda\rho}
{\cal F}_{\rho\sigma}{\cal F}^{\sigma}_{\;\;\nu} \right), \n
&&\pi_{\phi}=\frac{1}{\sqrt{-g}}\frac{\delta S_0}{\delta \phi}
=-2\alpha e^{-2\phi+\rho} -\beta e^{-\phi} A , \n
&&\pi_{\rho}=\frac{1}{\sqrt{-g}}\frac{\delta S_0}{\delta \rho}
=\alpha e^{-2\phi+\rho}, \n
&&\pi_{B\mu\nu}=g_{\mu\lambda}g_{\nu\rho}
\frac{1}{\sqrt{-g}}\frac{\delta S_0}{\delta B_{\lambda\rho}} \n
&&\qquad \;\: =\frac{\beta e^{-\phi}}{2A} \left( {\cal F}_{\mu\nu}
+\frac{1}{2}{\cal F}_{\mu\nu}{\cal F}_{\lambda\rho}{\cal F}^{\lambda\rho}
+{\cal F}_{\mu\lambda}{\cal F}^{\lambda\rho}{\cal F}_{\rho\nu} \right)
+\frac{\gamma}{4}\varepsilon_{\mu\nu}^{\;\;\;\;\lambda\rho}
(C_{\lambda\rho}+\chi {\cal F}_{\lambda\rho}), \n
&&\pi_{\chi}=\frac{1}{\sqrt{-g}}\frac{\delta S_0}{\delta \chi}
=\frac{\gamma}{8}
\varepsilon^{\mu\nu\lambda\rho}{\cal F}_{\mu\nu}{\cal F}_{\lambda\rho}, \n
&&\pi_{C\mu\nu}=g_{\mu\lambda}g_{\nu\rho}
\frac{1}{\sqrt{-g}}\frac{\delta S_0}{\delta C_{\lambda\rho}}
=\frac{\gamma}{4}
\varepsilon_{\mu\nu}^{\;\;\;\;\lambda\rho}{\cal F}_{\lambda\rho}, \n
&&\pi_{D\mu\nu\lambda\rho}
=g_{\mu\mu'}g_{\nu\nu'}g_{\lambda\lambda'}g_{\rho\rho'}
\frac{1}{\sqrt{-g}}\frac{\delta S_0}{\delta D_{\mu'\nu'\lambda'\rho'}}
=\frac{\gamma}{24}\varepsilon_{\mu\nu\lambda\rho},
\label{pi}
\eeqa
where
\beqa
A \equiv \sqrt{\frac{\det ({\cal G}+{\cal B})}{\det {\cal G}}}
=\sqrt{1+\frac{1}{2}{\cal F}_{\mu\nu}{\cal F}^{\mu\nu}
+\frac{1}{8}({\cal F}_{\mu\nu}{\cal F}^{\mu\nu})^2
-\frac{1}{4}{\cal F}_{\mu\nu}{\cal F}^{\nu\lambda}
{\cal F}_{\lambda\rho}{\cal F}^{\rho\mu}}.
\nonumber
\eeqa
We substitute the values of the fields in (\ref{D3branesolution}) and
(\ref{background}) into
the right-hand sides of (\ref{pi}), setting
\beqa
F_{\mu\nu}=0.
\nonumber
\eeqa
Noting that $A=h_0^{-1/2}/\cos\theta$
for (\ref{D3branesolution}) and $A=\sqrt{1+a^4u_0^4}$ for (\ref{background})
on the boundaries,
it can be easily verified that
the right-hand sides of (\ref{pi}) reproduce the right-hand sides of 
(\ref{onshellpi1}) if 
\beq
\alpha=10 \;\;\; \mbox{and} \;\;\; 
\beta=-\gamma=-\frac{4{\alpha'}^2R^4\cos\theta}{g},
\label{alphabeta1}
\eeq
and the right-hands of (\ref{onshellpi2}) if
\beq
\alpha=10 \;\;\; \mbox{and} \;\;\; 
\beta=-\gamma=-\frac{4{\alpha'}^2R^4}{\hat{g}}.
\label{alphabeta2}
\eeq
These conditions are consistent with (\ref{condition}), and (\ref{alphabeta1})
reduces to (\ref{alphabeta2}) in the near-horizon limit (\ref{scaling}).

Next, we compare the value of the on-shell action with that of $S_0$ directly.
We substitute the values of the fields with $r=r_0$ in (\ref{D3branesolution})
and the values of the fields with $u=u_0$ in (\ref{background})
into (\ref{ScSBISWZ}), respectively. 
For (\ref{D3branesolution}), we obtain
\beqa
&&S_c=\frac{\alpha V_4 r_0^4}{g^2}, \n
&&S_{BI}=\beta \int d^4x \; \sqrt{-g}\:e^{-\phi}\:A
=\frac{\beta V_4}{g\cos\theta}f_0^{-1}, \n
&&S_{WZ}=\gamma \int d^4x \; (D_{0123}+C_{01}\:B_{23})
=\frac{\gamma V_4}{g\cos\theta}f_0^{-1},
\label{S0value1}
\eeqa
where $V_4=\int d^4x$.
For (\ref{background}), we obtain
\beqa
&&S_c=\frac{\alpha V_4 {\alpha'}^4R^8}{\hat{g}^2}u_0^4, \;\;\;
S_{BI}=\frac{\beta V_4 {\alpha'}^2R^4}{\hat{g}}u_0^4, \;\;\;
S_{WZ}=\frac{\gamma V_4 {\alpha'}^2R^4}{\hat{g}}u_0^4.
\label{S0value2}
\eeqa
Here, we set
\beqa
\sigma=0.
\nonumber
\eeqa
Then, it follows from (\ref{S0}), (\ref{alphabeta1}), (\ref{alphabeta2}), 
(\ref{S0value1}) and (\ref{S0value2}) that
\beqa
S_0=S_c+S_{BI}+S_{WZ}=\left\{ \begin{array}{ll}
\frac{10 V_4}{g^2}r_0^4  & 
\mbox{for (\ref{D3branesolution})} \\
\frac{10 V_4 {\alpha'}^4R^8}{\hat{g}^2}u_0^4  
& \mbox{for (\ref{background})}.
\end{array} \right.
\label{S0value}
\eeqa
In (\ref{S0value1}) and (\ref{S0value}), the quantities for 
(\ref{D3branesolution}) reduce to those for (\ref{background})
in the near-horizon limit (\ref{scaling}).
We calculate the values of the on-shell actions for (\ref{D3branesolution})
and (\ref{background}) by substituting
(\ref{onshellpi1}) by $r_0$ replaced by $r$ and (\ref{onshellpi2})
with $u_0$ replaced by $u$ into (\ref{ADM}), respectively. Noting
that the constraints in (\ref{ADM}) are satisfied on shell, we 
reproduce the value of $S_0$ for (\ref{D3branesolution}) as follows:
\beqa
I_5^{on-shell}&=&\int_0^{r_0} dr d^4x \;
\sqrt{-g}(\pi^{\mu\nu}\pa_r g_{\mu\nu}+\pi_{\phi}\pa_r \phi
+\pi_{\rho}\pa_r \rho+\pi_B^{\mu\nu} \pa_r B_{\mu\nu} \n
&&\qquad \qquad \qquad  \;\;\;\;+\pi_{\chi}\pa_r \chi
+\pi_C^{\mu\nu} \pa_r C_{\mu\nu} 
+\pi_D^{\mu\nu\lambda\rho} \pa_r D_{\mu\nu\lambda\rho}) \n
&=&\frac{40 V_4}{g^2} \int_0^{r_0} dr \; r^3 \n
&=&\frac{10 V_4}{g^2} r_0^4.
\eeqa
We reproduce the value of $S_0$ for (\ref{background}) in the same way.
Thus, we have shown that the on-shell actions
for the supergravity solution (\ref{D3branesolution}) 
and its near-horizon limit
(\ref{background}) are reproduced by our solution (\ref{S0}) with
$F_{\mu\nu}=0$ and $\sigma=0$ when $\alpha$ and $\beta$ take the values in
(\ref{alphabeta1}) and (\ref{alphabeta2}), respectively.

\vspace*{1cm}

\section{Effective action of a probe D3-brane}
\setcounter{equation}{0}
In this section, we show that the solution (\ref{S0}) to the H-J equation 
obtained in section 4 is the effective action of a probe D3-brane. 
In section 4, we obtained $S_0$ as a solution 
to the H-J equation (\ref{HJ}). $S_0$ is a functional of
the boundary values of the fields and an on-shell action for a set of 
solutions of the five-dimensional gravity (\ref{5Daction}).
In section 5, we showed that the supergravity solutions 
(\ref{D3branesolution}) and (\ref{background}) belong to this set.
Intuitively, the solution to the H-J equation
corresponds to the effective action of a probe D3-brane located
inside and outside the near-horizon
region of a stack of D3-branes, and $u$ and $r$ correspond to the position 
of the probe D3-brane. In what follows, we give arguments that justify this 
interpretation.

The quantities $S_{BI}$ and $S_{WZ}$ in (\ref{S0}) together take the form of 
the D3-brane effective action
if the metric, the dilaton, the anti-symmetric field and the R-R fields
in those terms can be regarded as those induced
in the D3-brane world-volume and $F_{\mu\nu}$ can be regarded as
the $U(1)$ gauge field strength.
Thus, if the above identifications are valid and $S_c$ can be ignored,
our interpretation is justified.
 
First, let us recall the relation between the fields in the target space and 
the induced fields in the world-volume.
Let $\zeta^{\mu}$ $(\mu=0,\cdots,3)$ be the coordinates of the 
D3-brane world-volume and 
$\bar{\xi}^{\alpha}(\zeta)$ $(\alpha=0,\cdots,4)$
be the embedding functions of the D3-brane 
in the five dimensions $\xi^{\alpha}$. 
The induced metric on the 
D3-brane is defined by the pull back
\beq
\bar{g}_{\mu\nu}(\zeta)=\frac{\pa \bar{\xi}^{\alpha}}{\pa \zeta^{\mu}}
\frac{\pa \bar{\xi}^{\beta}}{\pa \zeta^{\nu}} \: h_{\alpha\beta}(\bar{\xi}).
\eeq
The other induced fields in the world-volume 
are defined by the pull back in the same way.
The effective action of the D3-brane is in general expressed in terms of these
induced fields.
When one considers the `flat' D3-brane, $\bar{\xi}^4(\zeta)$ is constant,
so that the induced fields are equivalent to the fields in the target space, 
up to a diffeomorphism in the world-volume. 
In our calculation, the fixed-time surface corresponds to the world-volume of
the probe D3-brane.
Hence, the above situation is realized, and
the static gauge 
$\zeta^{\mu}=\bar{\xi}^{\mu}\:(=x^{\mu})$ is adopted in our calculation,
so that the induced fields coincide with the original 
fields in the target space.

Next, let us show that $F_{\mu\nu}$ in the solution to the H-J equation
corresponds to
the $U(1)$ gauge field strength.
One can see that (\ref{D3branesolution})
and (\ref{background}) remain solutions of the five-dimensional
gravity even if the fields are deformed as
\beqa
B_{\mu\nu} &\rightarrow& B_{\mu\nu}+b_{\mu\nu}, \n
C_{\mu\nu} &\rightarrow& C_{\mu\nu}+c_{\mu\nu}, \n
D_{\mu\nu\lambda\rho} &\rightarrow& D_{\mu\nu\lambda\rho}
+d_{\mu\nu\lambda\rho}-6 \; c_{[\mu\nu}B_{\lambda\rho]},
\label{transformation}
\eeqa
where $b_{\mu\nu}$, $c_{\mu\nu}$ and $d_{\mu\nu\lambda\rho}$ are constants.
In fact, if we introduce $\Lambda$, $\Sigma$ and $\Xi$ in such a way that
\beqa
b_{\mu\nu} &=& \partial_{\mu}\Lambda_{\nu}-\partial_{\nu}\Lambda_{\mu}, \n
c_{\mu\nu} &=& \partial_{\mu}\Sigma_{\nu}-\partial_{\nu}\Sigma_{\mu}, \n
d_{\mu\nu\lambda\rho} &=& 4 \; \partial_{[\mu} \Xi_{\nu\lambda\rho]},
\label{bcd}
\eeqa
the transformations (\ref{transformation}) are identical to 
the transformations for the $U(1)$ gauge symmetries in type IIB supergravity, 
as explained in appendix C. However, $S_0$ does not  need to be invariant
under (\ref{transformation}), because the partial integrations in
the arguments in appendix C fail
for $\Lambda$, $\Sigma$ and $\Xi$ in (\ref{bcd}). Instead, $F_{\mu\nu}$ and
$\sigma$ in $S_0$ are transformed under (\ref{transformation}) as
\beqa
F_{\mu\nu} &\rightarrow& F_{\mu\nu}-b_{\mu\nu}, \n
\sigma &\rightarrow& \sigma - \gamma \; V_4 \; (d_{0123}+6 \; c_{[01}F_{23]} );
\eeqa
that is, $F_{\mu\nu}$ is equivalent with the constant shift of 
$B_{\mu\nu}$ caused by the gauge transformation. Therefore, it follows 
from a standard argument in string theory that $F_{\mu\nu}$ should be
the $U(1)$ gauge field strength in the D3-brane world-volume.

Finally, in order to justify dropping $S_c$, let us see 
the dependence of each term in (\ref{S0}) on the dilaton field.
By redefining the R-R fields in such a way that the action of type IIB
supergravity is multiplied by an overall factor of $e^{-2\Phi}$, one can see
that $S_{BI}$ and $S_{WZ}$ are proportional to $e^{-\phi}$. As is well
known, this fact indicates that these terms come from the disk diagram 
in string theory. On the other hand, $S_c$ is proportional to $e^{-2\phi}$.
It is natural to consider $S_c$ to come from the sphere diagram 
in string theory, which corresponds to (the logarithm of)
the vacuum transition amplitude or the vacuum bubble diagram.
Therefore $S_c$ should be subtracted from the contribution to the effective
action of the probe D3-brane.
Thus $S_{BI}+S_{WZ}$ in the solution to the H-J equation is interpreted as 
the effective action of the 
probe D3-brane.

\vspace{1cm}

\section{Summary and discussion}
\setcounter{equation}{0}
In this paper, we showed that the D3-brane effective action plus the 
cosmological term is a solution to
the H-J equation in type IIB supergravity. 
This solution to the H-J equation reproduces the on-shell
actions for the near-horizon geometries of a stack of D-branes and should
correspond to the effective action in the dual Yang Mills with a nontrivial
vacuum expectation value of the Higgs field. It also reproduces the on-shell
action for the supergravity solution of a stack of D3-brane in a $B_2$ field 
without the near-horizon limit.
Obtaining an interpretation of this result is an open problem. 
Our findings are expected to
be a prototype for the calculations through which the correspondence
between gauge theories and gravities is checked. They should also
shed light on the holographic
renormalization group flow generated by the perturbation of the operators
dual to the tensor fields as well as on the holographic renormalization (group)
in noncommutative Yang Mills.

We can apply similar calculations to the case of general D$p$-branes.
In fact, we have already verified that the D$p$-brane effective action plus
a cosmological term analogous to $S_c$ form a solution to the H-J
equation in type IIA(IIB) supergravity reduced on $S^{8-p}$  for $p=1,\:2$.
This result is natural from the viewpoint of Ref.\cite{JKY}, where it is shown
that the effective action of a probe D$p$-brane in the near-horizon
geometry generated by a stack of D$p$-branes, which takes the form of 
the Born-Infeld action, is determined only by the generalized conformal 
symmetry. 
It is relevant to investigate whether these solutions to the H-J equations
reproduce the on-shell actions for the supergravity solutions of 
the D$p$-brane 
corresponding to noncommutative
Yang Mills in $p+1$ dimensions and our results can be extended to the cases
$p > 3$. We hope to report studies of these problems in the near future.

Some comments are in order.
In this paper, we solved the H-J equation under the ansatz that
the fields are constant on the fixed-time surface. In other words, 
we solved the equation in the mini-superspace approximation. 
Our solution to the H-J equation corresponds to
the lowest terms in the derivative expansion 
of the on-shell action \cite{dBVV}, whose derivatives with respect to the
fields give the beta functions and the anomalous dimensions.
It is important to obtain the higher-order
terms in the derivative expansion, going beyond this ansatz, and
elucidate what in the Coulomb branch of the dual Yang Mills  
corresponds to these higher 
order terms. One can also study the structure of 
the holographic renormalization group in ordinary and 
noncommutative Yang Mills in terms of the higher-order terms.
We considered only the `flat' D-branes in this paper. In order to treat
a D-brane fluctuating in the directions transverse to
the world-volume, we have to develop a new formalism that generalizes the
ADM formalism in the gravitational system and enables us to consider a `local'
time. 

\vspace*{1cm}

\section*{Acknowledgements}
A.T. would like to thank M. Fukuma, W. Taylor, J. Troost and T. Yoneya 
for stimulating discussions. M.S. would like to thank H. Emoto for stimulating
discussions.  We are also grateful to
Y. Hosotani and D. Tomino for useful conversations. 
The work of A.T. is supported in part by funds provided by the U.S.
Department of Energy (D.O.E.) under cooperative research agreement
DF-FC02-94ER40818, while the work of M.S. is supported in part 
by Research Fellowships of the Japan 
Society for the Promotion of Science (JSPS) for Young Scientists (No.13-01193).

\vspace*{1cm}

\section*{Appendix A: Equations of motion in type IIB supergravity }
\setcounter{equation}{0}
\renewcommand{\theequation}{A.\arabic{equation}}
In this appendix, we list explicitly
the equations of motion and the constraints
for type IIB supergravity. We have
\beqa
&&R_{MN}^G+2D_M D_N \Phi 
-\frac{1}{4}H^{(3)}_{M L_1L_2}H_N^{(3)L_1L_2}
-\frac{1}{2}e^{2\Phi}F^{(1)}_M F^{(1)}_N
-\frac{1}{4}e^{2\Phi}\tilde{F}^{(3)}_{M L_1L_2}\tilde{F}_N^{(3)L_1L_2} \n
&&-\frac{1}{4\cdot 4!}e^{2\Phi}\tilde{F}^{(5)}_{M L_1 \cdots L_4}
\tilde{F}_N^{(5)L_1 \cdots L_4} \n
&&+G_{MN}\left(-\frac{1}{2}R_G -2D_L D^L \Phi
+2\pa_L \Phi \pa^L \Phi +\frac{1}{4}(|H_3|^2
+e^{2\Phi}|F_1|^2 +e^{2\Phi}|\tilde{F}_3|^2) \right) \n
&&=0, \\
&&R_G +4D_M D^M \Phi -4\pa_M \Phi \pa^M \Phi -\frac{1}{2}|H_3|^2=0, \\
&&D_L(e^{-2\Phi}H^{LMN})+D_L(C_0 \tilde{F}^{(3)LMN}) 
+\frac{1}{6}F^{(3)}_{L_1 L_2 L_3} \tilde{F}^{(5)MN L_1 L_2 L_3}=0, \\
&&D_L F^{(1)L} -\frac{1}{6}H^{(3)}_{L_1 L_2 L_3}\tilde{F}^{(3)L_1 L_2 L_3}=0,\\
&&D_L \tilde{F}^{(3)LMN} -\frac{1}{6}H^{(3)}_{L_1 L_2 L_3}
\tilde{F}^{(5)MN L_1 L_2 L_3}=0, \\ 
&&D_L \tilde{F}^{(5)L M_1 M_2 M_3 M_4} 
+\frac{1}{36}\varepsilon^{M_1 M_2 M_3 M_4 L_1 \cdots L_6} 
H^{(3)}_{L_1 L_2 L_3} F^{(3)}_{L_4 L_5 L_6} = 0,
\eeqa
where $D_M$ represents the covariant derivative in ten dimensions.
The self-duality condition for the five-form (\ref{self-duality})
is expressed explicitly as
\beq
\tilde{F}^{(5)M_1 M_2 M_3 M_4 M_5}= 
\frac{1}{5!} \varepsilon ^{M_1 M_2 M_3 M_4 M_5 L_1 \cdots L_5} 
\tilde{F}^{(5)}_{L_1 L_2 L_3 L_4 L_5}.
\eeq

\vspace{1cm}

\section*{Appendix B: Some useful formulae}
\setcounter{equation}{0}
\renewcommand{\theequation}{B.\arabic{equation}}
Let us consider the following reduction of the ten-dimensional space-time 
on $S^{8-p}$:
\beqa
ds_{10}^{\:2}&=&G_{MN} \: dX^M dX^N \n
&=& h_{\alpha\beta}(\xi) \: d\xi^{\alpha}d\xi^{\beta} 
+e^{\rho(\xi)/2} \: d\Omega_{8-p}.
\eeqa
Here, the $\xi^{\alpha}$ are $(p+2)$-dimensional coordinates, and $S^{8-p}$ is
parametrized by $\theta_1,\cdots,\theta_{8-p}$. The ten-dimensional curvatures
are represented by the $(p+2)$-dimensional curvatures
and the $(8-p)$-dimensional
curvatures as
\beqa
&&R^G_{\alpha\beta}=R^{(p+2)}_{\alpha\beta}
-\frac{8-p}{4}\left(\nabla^{(p+2)}_{\alpha}\nabla^{(p+2)}_{\beta}\rho
+\frac{1}{4}\pa_{\alpha}\rho \: \pa_{\beta}\rho \right), \n
&&R^G_{\theta_i \theta_j}=R^{(S^{8-p})}_{\theta_i \theta_j}
+\left( -\frac{1}{4}\nabla^{(p+2)}_{\alpha}\nabla^{(p+2)\alpha}\rho
-\frac{8-p}{16}\pa_{\alpha}\rho \: \pa^{\alpha}\rho \right) \: e^{\rho/2} \:
g^{(S^{8-p})}_{\theta_i \theta_j}, \n
&&R_G=R^{(p+2)}-\frac{8-p}{2}\nabla^{(p+2)}_{\alpha}\nabla^{(p+2)\alpha}\rho
-\frac{(8-p)(9-p)}{16}\pa_{\alpha}\rho \: \pa^{\alpha}\rho
+e^{-\rho/2} \: R^{(S^{8-p})},
\eeqa
where $R^{(S^{8-p})}$ is the constant curvature of $S^{8-p}$.

\vspace*{1cm}

\section*{Appendix C: Momentum constraint and Gauss law constrains}
\setcounter{equation}{0}
\renewcommand{\theequation}{C.\arabic{equation}}
In this appendix, we elucidate the momentum constraint and the
Gauss law constraints. First, note that the five-dimensional action
(\ref{5Daction})
is invariant under the following $U(1)$ transformations:
\beqa
\delta_{gauge} B &=& d \Lambda, \n
\delta_{gauge} C &=& d \Sigma, \n
\delta_{gauge} D &=& d \Xi - \Sigma \wedge H_3.
\label{U(1)}
\eeqa
The Gauss law constraints $Z_B=0$, $Z_C=0$ and $Z_D=0$ imply that the following
relations hold for arbitrary $\Lambda$, $\Sigma$ and $\Xi$, respectively:
\beqa
0&=&\int d^4 x \sqrt{-g} (\pa_{\mu}\Lambda_{\nu}-\pa_{\nu}\Lambda_{\mu})
\frac{1}{\sqrt{-g}}\frac{\delta S}{\delta B_{\mu\nu}}, \n
0&=&\int d^4 x \sqrt{-g} \left( (\pa_{\mu}\Sigma_{\nu}-\pa_{\nu}\Sigma_{\mu})
\frac{1}{\sqrt{-g}}\frac{\delta S}{\delta C_{\mu\nu}} \right. \n
&&\qquad \qquad \qquad \left. 
-(\Sigma_{\mu}H_{\nu\lambda\rho}-\Sigma_{\nu}H_{\lambda\rho\mu}
+\Sigma_{\lambda}H_{\rho\mu\nu}-\Sigma_{\rho}H_{\mu\nu\lambda})
\frac{\delta S}{\delta D_{\mu\nu\lambda\rho}} \right), \n
0&=&\int d^4 x \sqrt{-g} \left( (\pa_{\mu}\Xi_{\nu\lambda\rho}
-\pa_{\nu}\Xi_{\lambda\rho\mu}+\pa_{\lambda}\Xi_{\rho\mu\nu}
-\pa_{\rho}\Xi_{\mu\nu\lambda})
\frac{1}{\sqrt{-g}}\frac{\delta S}{\delta D_{\mu\nu\lambda\rho}} \right).
\label{Gausslawconstraints}
\eeqa 
These relations  indicate that
$S$ is invariant under the $U(1)$ gauge transformations on the
fixed-time surface. 
Finally, we examine the momentum constraint 
$H^{\mu}=0$. For an arbitrary infinitesimal parameter $\varepsilon^{\mu}$,
this constraint leads to the relation
\beqa
0&=&\int d^4 x \sqrt{-g} \varepsilon_{\mu} \left(
-2\nabla_{\nu} \left(\frac{1}{\sqrt{-g}}\frac{\delta S}{\delta g_{\mu\nu}}
\right) +\pa^{\mu}\phi \frac{1}{\sqrt{-g}}\frac{\delta S}{\delta \phi}
+\pa^{\mu} \rho \frac{1}{\sqrt{-g}}\frac{\delta S}{\delta \rho}
+H^{\mu}_{\;\;\nu\lambda} 
\frac{1}{\sqrt{-g}}\frac{\delta S}{\delta B_{\nu\lambda}} \right. \n
&&\qquad \qquad \qquad 
\left. +\pa^{\mu}\chi \frac{1}{\sqrt{-g}}\frac{\delta S}{\delta \chi}
+F^{\mu}_{\;\;\nu\lambda} 
\frac{1}{\sqrt{-g}}\frac{\delta S}{\delta C_{\nu\lambda}}
+(G^{\mu}_{\;\;\nu\lambda\rho\sigma}+4C^{\mu}_{\;\;\nu}H_{\lambda\rho\sigma})
\frac{1}{\sqrt{-g}}\frac{\delta S}{\delta D_{\nu\lambda\rho\sigma}}
\right) \n
&=&\int d^4 x \sqrt{-g} \left(
\delta_{diff} g_{\mu\nu} \frac{1}{\sqrt{-g}}\frac{\delta S}{\delta g_{\mu\nu}}
+\delta_{diff} \phi \frac{1}{\sqrt{-g}}\frac{\delta S}{\delta \phi}
+\delta_{diff} \rho \frac{1}{\sqrt{-g}}\frac{\delta S}{\delta \rho}
+\delta_{diff} \chi \frac{1}{\sqrt{-g}}\frac{\delta S}{\delta \chi}
\right. \n
&&\qquad\qquad\qquad 
+(\delta_{diff} B_{\mu\nu} +\pa_{\mu}(\varepsilon^{\lambda}B_{\nu\lambda})
-\pa_{\nu}(\varepsilon^{\lambda}B_{\mu\lambda})) 
\frac{1}{\sqrt{-g}}\frac{\delta S}{\delta B_{\mu\nu}}\n
&&\qquad\qquad\qquad 
+(\delta_{diff} C_{\mu\nu} +\pa_{\mu}(\varepsilon^{\lambda}C_{\nu\lambda})
-\pa_{\nu}(\varepsilon^{\lambda}C_{\mu\lambda}))
\frac{1}{\sqrt{-g}}\frac{\delta S}{\delta C_{\mu\nu}} \n
&&\qquad\qquad\qquad
+(\delta_{diff} D_{\mu\nu\lambda\rho} 
+\pa_{\mu}(\varepsilon^{\sigma}D_{\nu\lambda\rho\sigma})
-\pa_{\nu}(\varepsilon^{\sigma}D_{\lambda\rho\mu\sigma})
+\pa_{\lambda}(\varepsilon^{\sigma}D_{\rho\mu\nu\sigma})
-\pa_{\rho}(\varepsilon^{\sigma}D_{\mu\nu\lambda\sigma}) \n
&&\qquad\qquad\qquad \left.
-\varepsilon^{\sigma}(C_{\mu\sigma}H_{\nu\lambda\rho}
-C_{\nu\sigma}H_{\lambda\rho\mu}+C_{\lambda\sigma}H_{\rho\mu\nu}
-C_{\rho\sigma}H_{\mu\nu\lambda}))
\frac{1}{\sqrt{-g}}\frac{\delta S}{\delta D_{\mu\nu\lambda\rho}} \right),
\label{diffeo}
\eeqa
where $\delta_{diff}$ stands for the diffeomorphism transformation with
respect to the parameter $\varepsilon^{\mu}$ 
on the fixed-time surface. By identifying 
$\varepsilon^{\sigma}B_{\mu\sigma}$, $\varepsilon^{\sigma}C_{\mu\sigma}$
and $\varepsilon^{\sigma}D_{\mu\nu\lambda\sigma}$ with $\Lambda_{\mu}$,
$\Sigma_{\mu}$ and $\Xi_{\mu\nu\lambda}$, respectively, one can see from
(\ref{Gausslawconstraints}) and (\ref{diffeo}) that $S$ must be
invariant under the diffeomorphism on the fixed-time surface.


\vspace*{1cm}
                        
\begin{thebibliography}{99}
\bibitem{Maldacena}J.M. Maldacena, Adv. Theor. Math. Phys. {\bf 2} (1998)  
231, hep-th/9711200.
\bibitem{GKP-W}S.S. Gubser, I.R. Klebanov and A.M. Polyakov,
Phys. Lett. {\bf B428} (1998) 105, hep-th/9802109. \\
E. Witten, Adv. Theor. Math. Phys. {\bf 2} (1998) 253,
hep-th/9802150.
\bibitem{DKPS}M.R. Douglas, D. Kabat, P. Pouliot and S.H. Shenker, 
Nucl. Phys. {\bf B485} (1997) 85, hep-th/9608024. 
\bibitem{AGMOO-DF}For a review, \\
O. Aharony, S.S. Gubser, J.M. Maldacena, H. Ooguri and Y. Oz, 
Phys. Rept. {\bf 323} (2000) 183, hep-th/9905111. \\
I.R. Klebanov, hep-th/0009139. \\
E. D'Hoker and D.Z. Freedman, hep-th/0201253.
\bibitem{DT}M.R. Douglas and W. Taylor, hep-th/9807225. 
\bibitem{Das}
A. Bilal and C. Chu, Nucl. Phys. {\bf B547} (1999) 179, hep-th/9810195. \\
P. Klaus, F. Larsen and S. Trivedi, JHEP {\bf 9903} (1999) 003,
hep-th/9811120. \\
S.R. Das, JHEP {\bf 9902} (1999) 012, hep-th/9901004; JHEP {\bf 9906} 029,
hep-th/9905037. 
\bibitem{Tseytlin}For a review, A.A. Tseytlin, {\it Born-Infeld action,
supersymmetry and string theory}, in the Yuri Golfand memorial volume: The
many faces of the superworld, ed. M. Shifman, World Scientific, hep-th/9908105.
\bibitem{BPT}I.L. Buchbinder, A.Yu. Petrov and A.A. Tseytlin, 
Nucl. Phys. {\bf B621} (2002) 179, hep-th/0110173.
\bibitem{HI-MR}A. Hashimoto and N. Itzhaki, Phys. Lett. {\bf B465} (1999) 142,
hep-th/9907166. \\
J.M. Maldacena and J.G. Russo, JHEP {\bf 9909} (1999) 25, hep-th/9908134.
\bibitem{Yoneya}T. Yoneya, Prog. Theor. Phys. Suppl. {\bf 134} (1999) 182,
hep-th/9902200.
\bibitem{dBVV}J. de Boer, E. Verlinde and H. Verlinde, JHEP {\bf 0008} (2000) 
003, hep-th/9912012.
\bibitem{FMS}M. Fukuma, S. Matsuura and T. Sakai, 
Prog. Theor. Phys. {\bf 104} (2000) 1089, hep-th/0007062;
Prog. Theor. Phys. {\bf 105} (2001) 1017, hep-th/0103187.
\bibitem{Corley}S. Corley, Phys. Lett. {\bf B484} (2000) 141, hep-th/0004030.
\bibitem{KM}J. Kalkkinen and D. Martelli, Nucl. Phys. {\bf B596} (2001) 415, 
hep-th/0007234.
\bibitem{NOZ}S. Nojiri, S.D. Odintsov and S. Zerbini, Phys. Rev. {\bf D62}
(2000) 064006, hep-th/0001192. 
\bibitem{NOO}S. Nojiri, S.D. Odintsov and S. Ogushi, 
Phys. Lett. {\bf B500} (2001) 199, hep-th/0011182. \\
S. Nojiri and S.D. Odintsov, Phys. Lett. {\bf B519} (2001) 145, hep-th/0106191.
\bibitem{MM}D. Martelli and W. Mueck, hep-th/0205061.
\bibitem{RT-BMM}J.G. Russo and A.A. Tseytlin, Nucl. Phys. {\bf B490} (1997) 121, hep-th/9611047. \\
J.C. Breckenridge, G. Michaud and R.C. Myers, Phys. Rev. {\bf D55} (1997) 6438, hep-th/9611174.
\bibitem{GH}G.W. Gibbons and S.W. Hawking, Phys. Rev. {\bf D15} (1977) 2752.
\bibitem{SW}
N. Seiberg and E. Witten, JHEP {\bf 9909} (1999) 32, hep-th/9908142.
\bibitem{JKY}A. Jevicki, Y. Kazama and T. Yoneya, 
Phys. Rev. {\bf D59} (1999) 066001, hep-th/9810146.

\end{thebibliography}
\end{document}